\documentclass[twocolumn,aps,showpacs,prb,tightenlines,amsmath,amssymb,superscriptaddress]{revtex4}
\usepackage{graphicx}
\usepackage{amssymb}
\usepackage{amsmath}
\usepackage{colordvi}
\begin{document}
\title{Spin-orbit coupling and $g$-factor of $X$-valley in cubic GaN}

\author{K. Shen}
\affiliation{Hefei National Laboratory for Physical Sciences
  at Microscale and Department of Physics, University of Science and Technology of
  China, Hefei, Anhui, 230026, China}
\author{J. Y. Fu}
\affiliation{Department of Physics, Qufu Normal University, Qufu, Shandong,
  273165, China}
\author{M. W. Wu}
\thanks{Author to whom correspondence should be addressed}
\email{mwwu@ustc.edu.cn.}
\affiliation{Hefei National Laboratory for Physical Sciences
  at Microscale and Department of Physics, University of Science and Technology of
  China, Hefei, Anhui, 230026, China}
\date{\today}
\begin{abstract}
 We report our theoretically investigation
on the spin-orbit coupling
and $g$-factor of the $X$-valley in cubic GaN. We find that the spin-orbit coupling
coefficient from $sp^3d^5s^\ast$ tight-binding model is 
0.029\,eV$\cdot${\AA}, which is comparable with that in cubic GaAs. By employing the
${\bf k}\cdot{\bf p}$ theory, we find that the $g$-factor in this case is only
slightly different from the free electron $g$-factor. These results are expected
to be important for the on-going study on spin dynamics far away from
equilibrium in cubic GaN.
\end{abstract}
% # 71.70.Ej Spin-orbit coupling, Zeeman and Stark splitting, Jahn-Teller effect
% # 85.75.?? d Magnetoelectronics; spintronics: devices exploiting spin polarized
%          transport or integrated magnetic fields
% # 61.82.Fk Semiconductors

\maketitle

%\section{Introduction}
Due to the existence of the wide energy gap between the conduction band and
the valence band, GaN has been proposed to be a promising candidate for
many electronic applications, such as the solid-state ultraviolet optical sources
and high-power electronic devices.\cite{naka} Recently, the discovery of the
room-temperature ferromagnetism in GaN based materials\cite{Dietl,lee}
highlights
its possible application in future spintronic devices.\cite{Zutic,Fabian,Awschalom,wu} Another
outstanding property of GaN for realizing the spintronic devices is the extremely
long spin lifetime,\cite{Weng} because of the relatively weak spin-orbit
coupling (SOC) compared to the narrow-gap III-V compounds, such as GaAs and
InAs.\cite{krish,Jiang}

For a detailed understanding of the spin dynamics,
the SOC is essential.\cite{wu}
In cubic GaN, the inequality of the cation and
anion in the crystal leads to the bulk inversion asymmetry, which results in
the Dresselhaus SOC.\cite{Dresselhaus}
Up to date, the investigation on spin
properties in GaN is focused on the low energy case, where only the lowest
valley, i.e., $\Gamma$-valley is relevant. Recently, Fu and Wu reported the
Dresselhaus SOC coefficient of the $\Gamma$-valley, 0.51~eV$\cdot${\AA}$^{3}$,
from the $sp^3s^\ast$ tight-binding (TB)
 model.\cite{Fu1} This result agrees with the later
experiment by the time-resolved Kerr rotation measurement.\cite{Bub1}
However, for the spin dynamics under the influence of
 high electric field\cite{conwell} 
or with spin pumping
by high-energy laser,\cite{Marie} electrons can be
driven into the high valleys.\cite{conwell} This multivalley correlation was
proposed to be able to induce the charge Gunn\cite{Gunn} and spin Gunn\cite{Qi}
effects in GaAs.  Zhang {\em et al.}\cite{Zhang} investigated the spin
dynamics under the high electric field in GaAs quantum wells and suggested that 
the spin Gunn effect
can be hindered by the fast spin relaxation 
of upper valley ($L$-valley) and hot-electron effect. To our best knowledge,
there has been no report on the multivalley spin dynamics in GaN till now. Since
the upper valley in GaN is the $X$-valley (no $L$-valley exists in cubic GaN),
one may expect different multivalley spin properties in this material. Therefore, the
details of the SOC in GaN is required.
In our previous work, the expression of the SOC of the $X$-valley in 
cubic III-V semiconductors was derived,\cite{Fu2} but the
corresponding coefficient of GaN is still unavailable. In the present communication, we calculate this coefficient for further investigations on the
spin dynamics in this material. Moreover, the $g$-factor
 is also required to take into account the
effect of the external magnetic field. 
Therefore, we will also calculate the $g$-factor of the
$X$-valley in GaN from the ${\bf k}\cdot {\bf p}$ theory.

%\section{Calculation and results}
%\subsection{Spin-orbit coupling}
In order to obtain the splitting energy of SOC, one needs to calculate the
band structure. One of the most widely used approaches for band structure
calculation is the TB theory.\cite{Jancu1} From our previous work on GaAs, we
find that the $d$-orbitals are important for the spin
splitting of the high valleys.\cite{Fu2,Diaz} Therefore, we employ the $sp^3d^5s^\ast$
nearest-neighbor TB model with the SOC here.\cite{Chadi} The parameters are
taken from the work by Jancu {\em et al.}.\cite{Jancu2}

\begin{figure}[bth]
  \centering  \includegraphics[width=8.0cm]{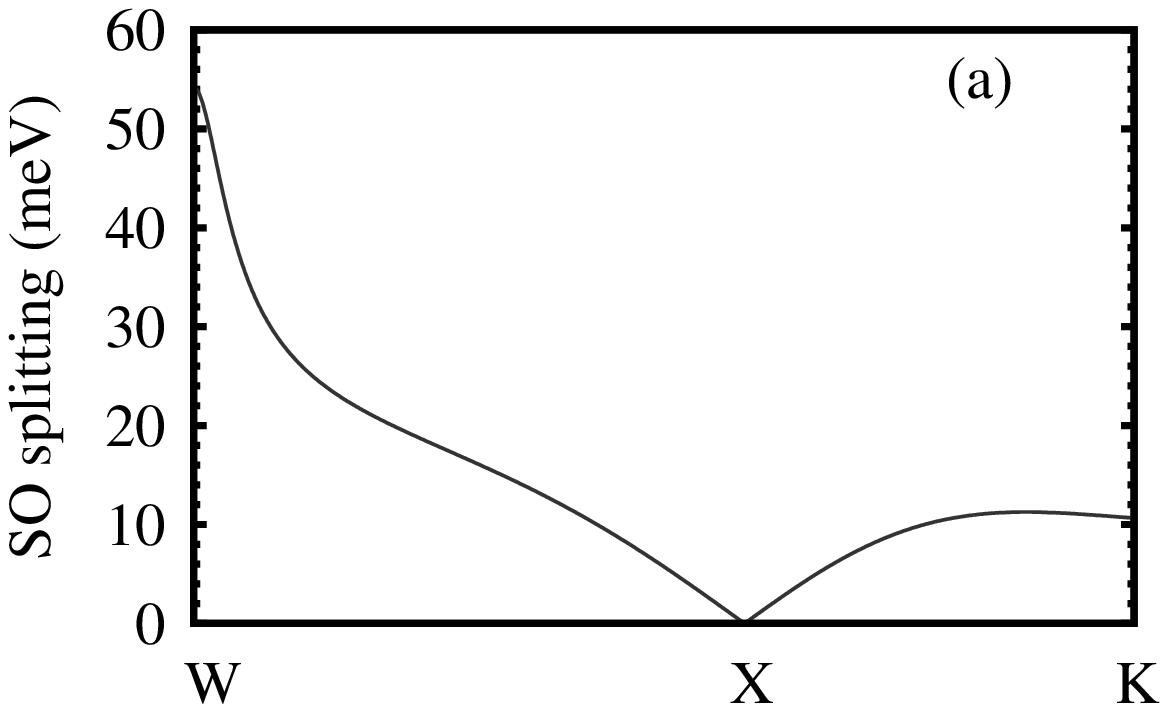}
  \par\vspace{-0.8cm}   
  \includegraphics[width=8.0cm]{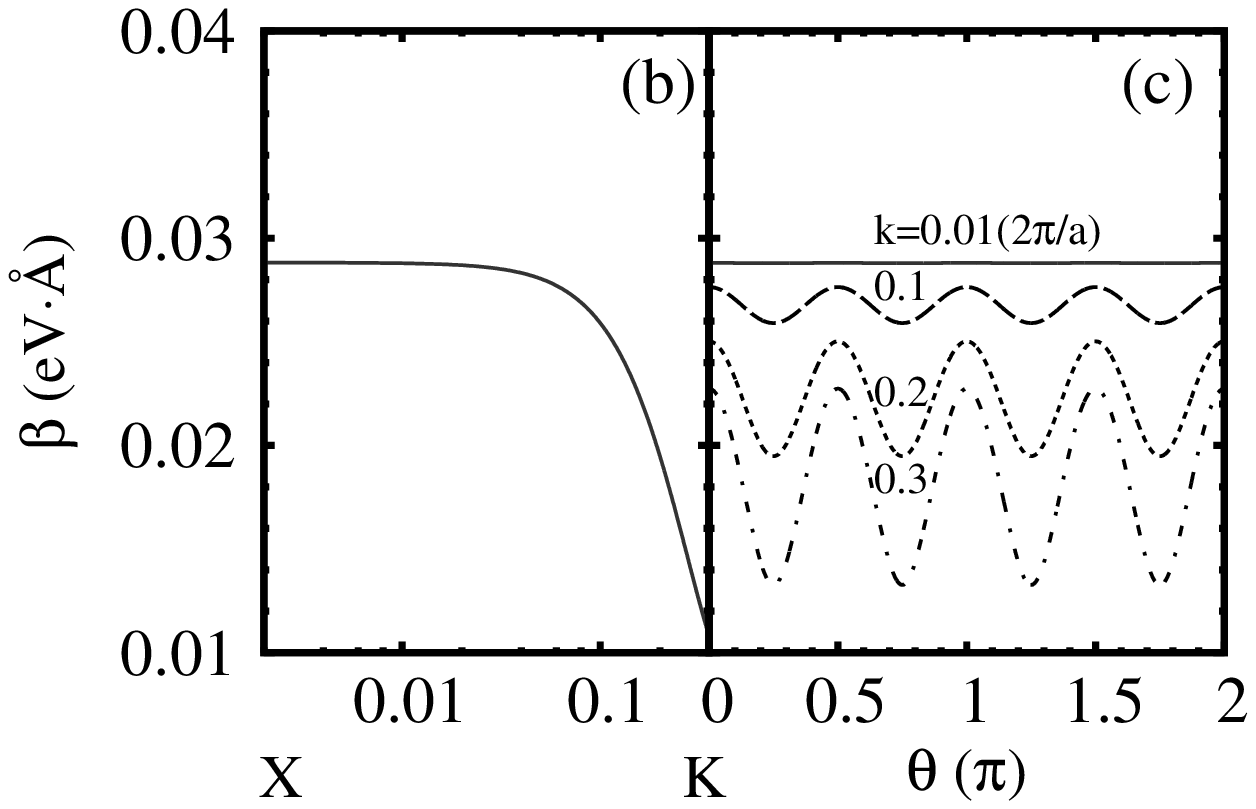}   
  \caption{(a) Spin-orbit splitting of cubic GaN
around the $X$-valley along $X \to K$ and $X \to W$
    directions. (b) 
    The corresponding SOC parameter $\beta$ vs. momentum along $X \to K$
    direction. (c) $\beta$ vs. angle for
      the momentum lying in the $x$-$y$ plane. Solid curve:
    $k$=0.01; dashed curve: 0.1; dotted curve: 0.2; and chain curve:
    0.3 (2$\pi$/$a$), with $a=4.5$~\AA, the lattice constant of cubic
GaN.\cite{vurga}
}
  \label{fig1}
\end{figure}

The spin-orbit splitting of the conduction band around the
$X$-valley is plotted as a
function of the momentum along $X \to K$ and $X \to W$ directions in
Fig.\,\ref{fig1}(a). One finds that the splitting increases
linearly with the momentum in the small momentum regime with respect to the
bottom of the $X$-valley. In the large momentum regime, this
monotonic tendency can be violated.

For the states close to the
$X$-point, the splitting can be described by the effective SOC
Hamiltonian, $\mathbf{\Omega}(\mathbf{k})
\cdot\mbox{\boldmath$\sigma$\unboldmath}$,
with $\mathbf{\Omega}$ and $\mbox{\boldmath$\sigma$\unboldmath}$ 
denoting the corresponding effective magnetic field of the conduction band
and the Pauli matrices.\cite{Ivchenko,Fu2} For the valleys lie in the [001]-direction, one
obtains\cite{Fu2} 
\begin{equation}  
\label{eq:x}  
\mathbf{\Omega}(\mathbf{k}) =\beta(k_x,-k_y,0).
\end{equation}
Here $\mathbf{k}$ represents the momentum measured
from the bottom of the valley. Obviously, this term results in the spin-orbit splitting
linearly depending on the momentum, i.e., $\Delta E=2\beta k_\|$ with
$k_\|=\sqrt{k^{2}_x+k^{2}_y}$  being the magnitude
of the transverse momentum. 

The SOC coefficient can be
measured by $\beta(\mathbf{k})=\Delta E/(2k_{\|})$. The value
of $\beta$ is shown as a function of momentum along $X \to K$ direction
in Fig.\,\ref{fig1}(b).
One can see that in the range of small momentum, $\beta$ keeps a
constant value 0.029\,eV$\cdot${\AA}. However, when the momentum lies far away 
from the bottom of the $X$-valley, $\beta$ decreases.
In the $X\to W$ direction, $\beta$
increases with increasing momentum (not shown) as expected from
Fig.\,\ref{fig1}(a).
This phenomenon is due to the higher order corrections of the SOC
terms. We now turn to the effect of
the direction of the transverse momentum on the SOC coefficient. In
Fig.\,\ref{fig1}(c), we show the anisotropic behavior of SOC
coefficient, where the momentum lies in the $x$-$y$ plane with $\theta$ being
the angle between the momentum and
$x$-axis. One can see the angle dependence becomes remarkable for large
momentum, which is also from the higher order correction of the SOC.
Interestingly, one notices that the SOC coefficient of
$X$-valley obtained here is comparable with that of GaAs,
0.059\,eV$\cdot${\AA},\cite{Fu2} which is 
very different from the situation of $\Gamma$-valley. For the $\Gamma$-valley,
the Dresselhaus SOC coefficient in GaN is much smaller than that in
GaAs. 

We should point out that the $d$-orbitals are of critical importance
in determining the SOC coefficient of the $X$-valley in GaN. Specifically, we obtain
$\beta=0.002$\,eV$\cdot${\AA} from $sp^3s^\ast$ TB model parameterized by
O'Reilly {\em et al.}, \cite{Reilly} which is one order of magnitude
smaller than that from $sp^3d^5s^\ast$ TB model. As a comparison, we also
calculate the SOC coefficient of the $\Gamma$-valley from $sp^3d^5s^\ast$ TB
model and obtain $\gamma=0.235$\,eV$\cdot${\AA}$^{3}$, which is close to that from
$sp^3s^\ast$ model $\gamma=0.508$\,eV$\cdot${\AA}$^{3}$ (Ref.\,\onlinecite{Fu1}).

%\subsection{$g$-factor}
Now, we turn to figure out the $g$-factor of the $X$-valley based on the ${\bf
  k}\cdot{\bf p}$ approach by following the approach given in
Ref.\,\onlinecite{shen}. In our calculation, we first include the conduction
band ($X_{1c}$) and the valence bands ($X_{3v}$ and $X_{5v}$) and neglect the
contribution of the other remote bands.

Similar to the $L$-valley case,\cite{shen} one can write the longitudinal effective
mass $m_l$ and the transverse one $m_t$ as\cite{shen} 
\begin{eqnarray}
  \frac{m_0}{m_l}-1={\frac{2}{m_0}}
  \frac{|\langle X_{1c}|p_z|X_{3v}\rangle|^2}{E_{X_{1c}}-E_{X_{3v}}},
  \label{eq2}
\end{eqnarray}
and
\begin{eqnarray}
  \frac{m_0}{m_t}-1=\frac{2}{m_0}
  \frac{|\langle X_{1c}|p_x|X_{5v}\rangle|^2}{E_{X_{1c}}-E_{X_{5v}}},
  \label{eq3}
\end{eqnarray}
respectively. 

The longitudinal and transverse $g$-factors are given by\cite{shen,roth}
\begin{eqnarray}
  \nonumber
  g_\|-g_0 &=& -\frac{2}{m_0}\frac{\delta
    \langle X_{1c}|p_x|X_{5v}\rangle\langle X_{5v}|p_y|X_{1c}\rangle
  }{(E_{X_{1c}}-E_{X_{3v}})(E_{X_{1c}}-E_{X_{5v}})}\\
  &=& -{\delta}(\tfrac{m_0}{m_t}-1)/({E_{X_{1c}}-E_{X_{3v}}}),
  \label{eq4}
\end{eqnarray}
and
\begin{eqnarray}
  \nonumber
  g_\perp-g_0&=&-{\frac{2}{m_0}}\frac{\delta^\prime
    \langle X_{1c}|p_y|X_{5v}\rangle\langle X_{3v}|p_z|X_{1c}\rangle
  }{(E_{X_{1c}}-E_{X_{3v}})(E_{X_{1c}}-E_{X_{5v}})}\\
  &=&-{\delta^\prime}\left[\frac{({m_0}/{m_t}-1)({m_0}/{m_l}-1)}
    {({E_{X_{1c}}-E_{X_{3v}}})({E_{X_{1c}}-E_{X_{5v}}})}\right]^{1/2},
  \label{eq5}
\end{eqnarray}
where $\delta=2i\langle X_{5v}|h_z|X_{5v}\rangle$ and 
$\delta^\prime=2i\langle X_{5v}|h_x|X_{3v}\rangle$ are the matrix elements of
the SOC.\cite{roth}  ${\bf p}$ describes the momentum
operator. $g_0$ is the $g$-factor of the free electron.

From our TB calculation, we obtain the energy levels of the $X$-valley, i.e.,
$E_{X_{1c}}=4.58$~eV, $E_{X_{5v}}=-2.74$~eV, and $E_{X_{3v}}=-6.98$~eV.
We estimate the matrix element of the SOC from the band splittings of the $X_{5v}$ band 
and take $\delta^\prime=\delta=0.02$~eV. With $m_l=0.5m_0$ and $m_t=0.3m_0$
(Ref.\,\onlinecite{vurga}), we
obtain $g_\|=1.996$ and $g_\perp=1.997$ by taking $g_0=2$. We should point out
that such a small
difference of the $g$-factor in $X$-valley from $g_0$ is consistent with the
previous results in silicon\cite{roth} and other III-V group compounds.\cite{baron}

Finally, we would like to discuss the contribution of the remote bands. One may notice
that the second conduction band ($X_{3c}$ with 
$E_{X_{3c}}=8.21$~eV) lies close to the $X_{1c}$
band.\cite{rubio} Since its symmetry is the same as that of the $X_{3v}$ band,
it can also contribute to 
$m_l$ and $g_\perp$. Similar to Eq.\,(\ref{eq2}), one calculates the
correction of the longitudinal effective mass due to the $X_{3c}$ band and finds
that the condition $\langle X_{1c}|p_z|X_{3c}\rangle/\langle X_{1c}|p_z|X_{3v}\rangle
<\sqrt{(E_{X_{3c}}-E_{X_{1c}})/(E_{X_{1c}}-E_{X_{3v}})}\approx 0.56$ 
to guarantee the real condition $m_l<m_0$. As a rough estimation, we take $\langle
X_{1c}|p_z|X_{3c}\rangle=0.5 \langle X_{1c}|p_z|X_{3v}\rangle$ and 
$\langle X_{5v}|h_x|X_{3c}\rangle=\langle X_{5v}|h_x|X_{3v}\rangle$. Then, it is
easy to calculate the final perpendicular $g$-factor $
g_\perp^\prime=g_0+(g_\perp-g_0)(1-0.5\eta)/\sqrt{1-0.25\eta}$ with
$\eta=(E_{X_{1c}}-E_{X_{3v}})/(E_{X_{3c}}-E_{X_{1c}})=3.18$ and
$g_\perp$ given by Eq.\,(\ref{eq5}). One obtains $g_\perp^\prime=2.004$, which
indicates that the relation
$g_\perp^\prime\approx g_0$ is still preserving by considering the correction from the
$X_{3c}$ band. 

%\section{Conclusion}
In summary, we have studied the SOC and $g$-factor of the $X$-valley in
cubic GaN. By taking into account the possible effect 
of the $d$-orbitals on high valleys, we perform our calculation with an
$sp^3d^5s^*$ nearest-neighbor TB model. The spin splitting of the conduction
band in the $X$-valley and the corresponding SOC coefficient are
calculated. We find that the SOC coefficient of the $X$-valley in GaN is
0.029\,eV$\cdot${\AA}, which is 
comparable with that in GaAs. In addition, we calculate the $g$-factor
and find that the value is very close to that of the free electron. These
results are useful for understanding the spin dynamics far away from 
the equilibrium.

%\begin{acknowledgments}
This work was supported by the Natural Science Foundation of China under Grants
No.~10725417. J.Y.F. acknowledges support from the Natural Science
Foundation of China (Grant No.~11004120) and the Youth Foundation of Qufu Normal
University.
%\end{acknowledgments}

\end{document}